# Flat Bands near Fermi Level of Topological Line Defects on Graphite

Lei Feng[1,§], Xianqing Lin[2,§], Lan Meng[1], Jia-Cai Nie[1], Jun Ni[2,*], and Lin He[1,*]

[1] Department of Physics, Beijing Normal University, Beijing, 100875, People's Republic of China

[2] Department of Physics and State Key Laboratory of Low-Dimensional Quantum Physics, Tsinghua University, Beijing, 100084, People's Republic of China

Flat bands play an important role in the study of strongly correlated phenomena, such as ferromagnetism, superconductivity, and fractional quantum Hall effect. Here we report direct experimental evidence for the presence of flat bands, close to the Fermi level, in one-dimensional topological defects on graphite seen as a pronounced peak in the tunnelling density of states. Our *ab initio* calculations indicate that the flat bands with vanishing Fermi velocity originate from $sp^2$ dangling bonds (with antibonding nature) of undercoordinated carbon atoms at the edges of the defects. We further demonstrate that the presence of flat bands could be a universal behavior of 1D defects of graphene/graphite with undercoordinated carbon atoms at the edges of the defects.

## I. INTRODUCTION

Graphene, in which carbon atoms are arranged in a two-dimensional (2D) honeycomb lattice, has ignited a tremendous outburst of scientific activities in both its fundamental physics and applications.[1-7] One of the most fascinating aspects of graphene is that its topological features of the electronic states can be fundamentally changed by modifying its local lattice structure.[8-13] Importantly, graphene's 2D nature makes it easier to add or remove carbon atoms to alter its electronic structures. This could provide a new set of building blocks and device concepts for an all-graphene circuit in the future. Very recently, Lahiri et al. observed a one-dimensional (1D) topological defect made of paired pentagons and octagons in an epitaxial graphene on a nickel substrate.[14] Their analysis based on density-functional theory (DFT) calculations predicted an almost flat band near Fermi energy ($E_F$), which makes the 1D defect have the potential to act as a conducting wire[14] and as a valley filter in the so-called valleytronics.[15,16] A flat band or Van Hove singularity near Fermi level is vital to many strongly correlated phenomena, such as ferromagnetism,[12,17] superconductivity,[18,19] and fractional quantum Hall effect.[20] Therefore, the flat band in 1D extended defect opens exciting opportunities for exploring correlated electronic phases in graphene. However, a direct experimental evidence for the presence of a flat band in 1D defect of graphene is still lacking so far.

In this paper, we present clear evidence for flat bands near Fermi energy in 1D extended defect on the surface of a highly oriented pyrolytic graphite (HOPG) by observing a pronounced peak in the tunnelling density of states. The 1D defects are tilt grain boundaries which are produced between two rotated topmost graphene grains. Our scanning tunneling microscopy (STM) measurements, complemented by DFT calculations, reveal well-localized electronic states near the Fermi energy around the topological line defects, which arise from flat bands with vanishing Fermi velocity. The flat bands originate from $sp^2$ dangling bonds (with antibonding nature) of undercoordinated carbon atoms at the edges of the defects. Our experimental result and analysis indicate that the presence of flat bands could be a universal behavior of similar 1D defects of graphene/graphite.

## II. EXPERIMENT

The STM system used in the experiments is an ultrahigh vacuum four-probe STM from UNISOKU. All the STM and scanning tunnelling spectroscopy (STS) measurements were carried out at liquid-nitrogen temperature in a constant-current scanning mode. The scaning tunnelling microscope tips were obtained from mechanically cut Pt (80%)/ Ir (20%) wire. All the STM images reported here were obtained by a tunnelling current ranging from 12.6 to 22.2 pA with sample bias voltage ranging from 438 to 600 mV unless otherwise specified. The dI/dV measurements were carried out with a standard lock-in technique using a 957 Hz a.c.

## III. RESULTS AND DISCUSSION

Samples of HOPG with the AB Bernal stacking, the most common in graphite, have been chosen to study in our experiments. HOPG samples were cleaved by an adhesive tape in air and transferred into the STM. Upon cleaving a HOPG sample, many different surface structures, such as strained graphene ridge[9] and grain boundaries,[21,22] can be observed in the resultant graphene layer.[23,24] The grain boundaries studied here are tilt grain boundaries which are produced between two rotated topmost graphene grains, as shown in Fig. 1. They usually show a 1D periodic superlattice with height corrugations arising from charge accumulation at the periodic defects.[23,24,25] Due to the charge accumulation, it is usually difficult to obtain atomic-resolution images along the 1D defect. Figure 2(a) shows a typical STM image of a 1D

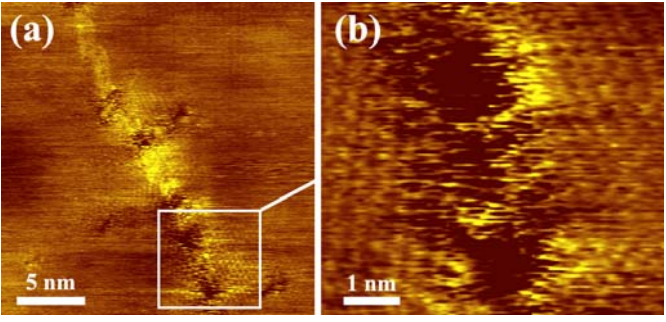

**FIG. 1** (color online). (a) STM image of a 1D defect with atomic resolution along the edge of the defect ($V_{sample}$ = 396 mV and I = 23.3 nA). (b) Zoom-in image of topography around the 1D defect.

grain boundary. The length of the 1D defect is about 1 μm and the periodicity along the line defect is about 2.0 nm. On the right hand of the 1D defect, another line structure with the same periodicity but with much smaller length (~ 50 nm) and height corrugations is observed. STS measurements along the left line defect show a pronounced peak near the Fermi level, as shown in Fig. 2(b). The standard V-shaped dI/dV curve of graphite is seen along the right line structure. It indicates that the right 1D defect may be a standing wave of electrons which reflects the topography of the left 1D defect.[26] The width, ~ 2.4 nm, of the left 1D defect indicates that carbon atoms (at least, partially carbon atoms) along the boundaries of the two rotated topmost graphene are undercoordinated. This could lead to creation of localized state at the Fermi energy.[27]

To explore the electronic properties of the line defect, spatially resolving STS spectra were measured. dI/dV vs bias voltage ($V_{bias}$) spectra on top of the 1D defect and of the flat graphite are shown in Figure 2(b). The tunnelling spectrum gives direct access to the local density of states (LDOS) of the surface at the position of the STM tip. While the spectrum in the flat region (inset) shows the standard V-shaped spectrum, a pronounced peak near the Fermi energy are clearly resolved in the spectrum measured on the 1D defect. Figure 2(c),(d) shows dI/dV spectra obtained at different positions (as marked) near the 1D defect shown in Fig. 2(a). The main features of the spectra are reproducible and the amplitude of the tunnelling peak near the Fermi level are almost independent of the positions along the 1D defect. The tunnelling peak, however, disappears completely when the STM tip is positioned slightly away from the 1D defect. As seen in the spectra of Fig. 2(d), the amplitude of the peak near the Fermi energy decreases as one moves away from the 1D defect, which indicates that the tunnelling peak arises from a quasi-localized state. The presence of flat bands near the Fermi level, which contributes a sharp peak to the DOS, in topological 1D defect of graphene is a fundamental result of this work. In literature, localized

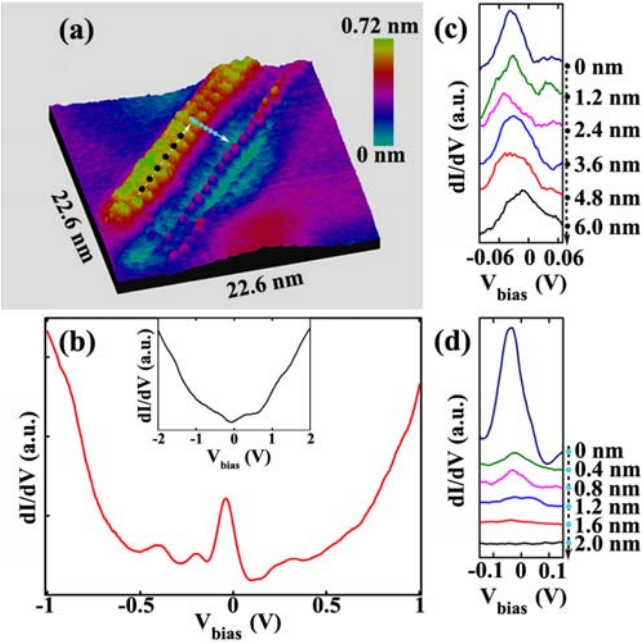

**FIG. 2** (color online). (a) 22.6 nm×22.6 nm constant-current STM image, measured at 77 K, of a line defect on HOPG surface ($V_{sample}$ = -600 mV and I = 12.6 nA). The other 1D structure on the right hand of the 1D defect could be the standing wave of electrons. (b) A typical dI/dV-V curve of the topological 1D defect measured at 77 K. A pronounced peak, close to the Fermi energy, is observed. The inset shows the dI/dV-V curve obtained on a flat graphite surface. (c) dI/dV spectra of the line defect shown in panel (a), measured at different points (black dots, as shown) along a line parallel to the 1D defect. The distance between these points is about 1.2 nm. (d) dI/dV spectra of the line defect shown in panel (a), measured at different points along a line perpendicular to the 1D defect. The distance between these points is about 0.6 nm.

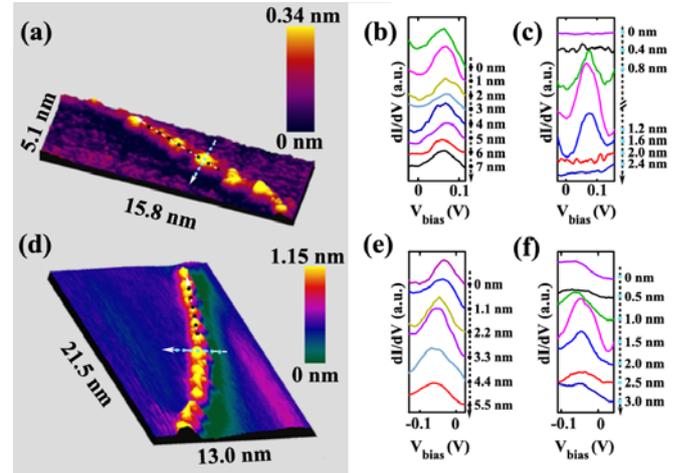

**FIG. 3** (color online). STM images and STS spectra of line defects on HOPG. (a) 5.1 nm ×15.8 nm constant-current STM image, measured at 77 K, of the second line defect on HOPG surface ($V_{sample}$ = 438 mV and I = 15.5 nA). The height of this line defect is about 0.4 nm and the width is about 1.2 nm. (b) dI/dV spectra of the line defect shown in panel (a), measured at different points along a line parallel to the 1D defect. The distance between these points is about 1 nm. (c) dI/dV spectra of the line defect shown in panel (a), measured at different points along a line perpendicular to the 1D defect. The distance between these points is about 0.4 nm. (d) 21.5 nm×13.0 nm constant-current STM image of the third line defect on HOPG surface ($V_{sample}$ = 600 mV and I = 22.2 nA). (e) dI/dV spectra of the line defect shown in panel (d), measured at different points along a line parallel to the 1D defect. The distance between these points is about 1.2 nm. (f) dI/dV spectra of the line defect shown in panel (d), measured at different point along a line perpendicular to the 1D defect. The distance between these points is about 0.5 nm.

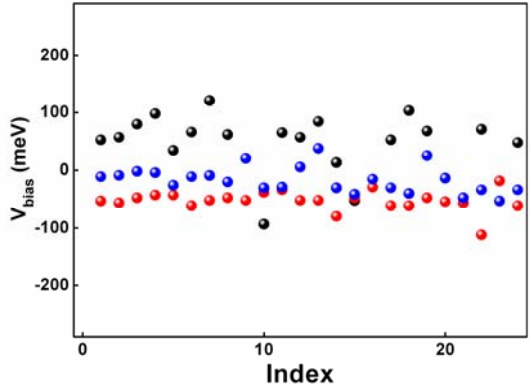

**FIG. 4** (color online). The position of the tunneling peak near the Fermi energy measured along the three 1D defects shown in Fig. 2 and Fig. 3. 25 dI/dV-V curves of each 1D defect are randomly selected to obtain the position of the tunneling peak. All of peak positions are within an energy interval of ±100 meV.

states have also been explored at point defect, steps, and aigzag edge of graphene/graphite systems.[12,28-32]

To further confirm the observed phenomena, we carried out STM and STS measurements systematically on another two 1D defects on HOPG surface. Figure 3 shows the STM images and the corresponding STS spectra. The periodicity along the line defects is about 3 nm. The spatially resolving STS spectra, as shown in Fig. 3(b),(c),(e),(f), show similar characteristics of that shown in Fig. 2. Figure 4 summarizes positions of the tunneling peak near the Fermi energy extracted from 25 dI/dV-V curves of the three 1D defects. The quasi-localized states near the Fermi level within an energy interval of ± 0.1 eV are observed along these 1D defects. Our experimental result indicates that the presence of flat bands could be a universal behavior of 1D defects of graphene/graphite.

We now turn to understand the microscopic origin of the flat bands of the 1D defects. The electronic properties of graphene with topological 1D defects have been studied by first-principles calculations. A unit cell of the real structure of the measured structures, as shown in Fig. 1-3, contains thousands atoms, making ab initio descriptions rather impractical. For simplicity, we use a structure of grain boundaries proposed by Cervenka et al[22] as our initial model, as shown in Fig. 5(a). The van der Waals coupling between graphene layers in graphite structure was not taken into account in the calculation. The proposed grain boundary consists of large vacancies with mostly zigzag edges. The misorientation angle between the grains is 17.9°. First-principles calculations have been performed to fully relax the structure and to study the peculiar electronic properties possessed by the grain boundary using the VASP code.[33] The Local Density Approximation (LDA)[34] is adopted with a kinetic energy cutoff of plane wave basis sets of 400 eV and the Vanderbilt ultrasoft pseudopotentials.[35] A rectangular supercell consisting of 196 carbon atoms is employed. The supercell has two parallel and equally spaced grain boundaries such that the periodic boundary condition is satisfied. The Brillouin zone sampling is done using a 4 ×

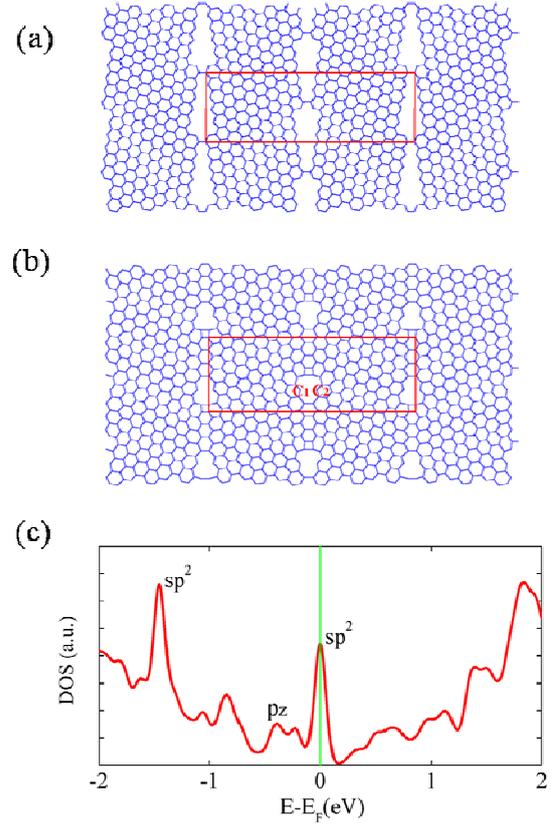

**FIG. 5** (color online). (a) The unrelaxed structure of the model system. (b) The relaxed structure of the model system. The solid line indicates the supercell involving tow grain boundaries. The carbon atoms with $sp^2$ dangling bonds in the vacancies are labeled as C1 and C2. (c) Density of states for the relaxed structure in panel (a). The peaks labeled as $sp^2$ and $p_z$ are contributed by the $sp^2$ dangling bonds and by the $p_z$ orbitals in the grain boundaries, respectively.

8 Monkhorst-Pack grid.[36] The vacuums in the direction perpendicular to the graphene plane are larger than 13 Å. The tolerance for the energy convergence is $10^{-5}$ eV. The structure is fully relaxed until the force on each atom is smaller than 0.01 eV/Å.

The relaxed structure is shown in Fig. 5(b). The vacancies of initial structure undergo great reconstruction as the $sp^2$ dangling bonds (DBs) of carbon atoms in different edges of the vacancies intend to form strong covalent bonds with each other when their distances are near enough. Covalent bonds form between atoms in different edges with distances smaller than 3.8 Å in the unrelaxed structure. Only the atoms labeled as C1 and C2 in the vacancies remain undercoordinated. The periodicity of the relaxed structure is 13.68 Å along the 1D defect. The calculated density of states (DOS) of the 1D defect is shown in Fig. 5(c). A sharp peak at the Fermi energy along with several small peaks around $E_F$ is observed. Obviously, the theoretical DOS around $E_F$ consists quite

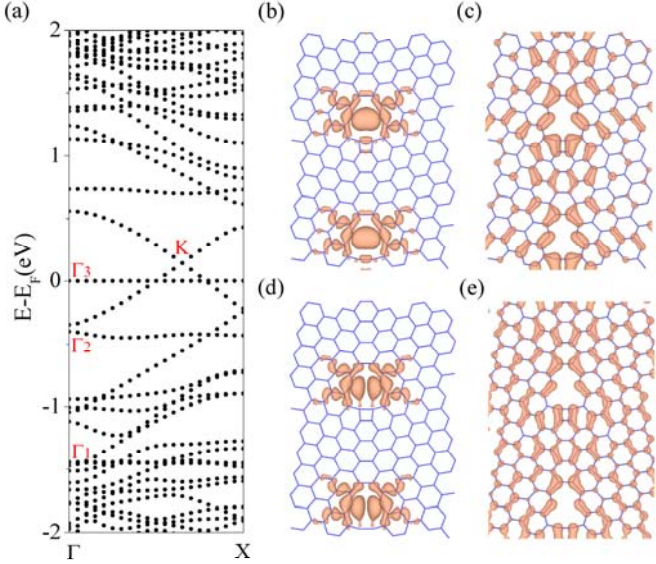

**FIG. 6** (color online). (a) The energy bands for the relaxed structure in panel (b) of Fig. 5. (b), (c), (d) and (e) show spatial distribution of the state $\Gamma_1$, $\Gamma_2$, $\Gamma_3$ and K labeled in panel (a), respectively.

well with our experimental result, as shown in Fig. 2(b). The DOS peaks around $E_F$ arise from contributions of the $sp^2$ dangling bonds and the $p_z$ orbitals in the 1D defects respectively, which will be elaborated subsequently.

Figure 6(a) shows the band structures along the k-point line parallel to the 1D defect (the k-point component perpendicular to the 1D defect is taken to be zero). There are two flat and almost degenerate bands lying near $E_F$. These bands are both empty. The spatial distribution of the state at the $\Gamma$ point in one of these bands is shown in Fig. 6(d) and is labeled as $\Gamma_3$ in Fig. 6(a). The $\Gamma_3$ state is localized around the vacancy and is composed of the DBs of C1 and C2 but with the antibonding characteristic. These two flat bands contribute to the sharp peak in DOS at $E_F$. The location of the flat bands with the antibonding nature between the DBs of C1 and C2 at $E_F$ can be attributed to the moderate distance between the undercoordinated atoms C1 and C2. The distance between C1 and C2 is 3.06 Å in the relaxed structure so that the interaction between the DBs of C1 and C2 are not strong enough to form the $\sigma$ covalent bond whose antibonding bands are far above $E_F$. The two bonding bands between the DBs of C1 and C2 appears 1.45 eV below $E_F$, labeled as $\Gamma_1$ in Fig. 6(a) also shows localized distribution around the vacancies, as shown in Fig. 6(b). In experiments, such DBs pairs in different edges with a moderate distance may be present in more complex vacancy configurations of 1D defects. It indicates that the presence of flat bands could be a universal behavior of 1D defects of graphene/graphite with undercoordinated carbon atoms at the edges of the defects.

Except the flat bands at $E_F$, the bands around $E_F$ all originate from the $p_z$ orbital network. The states in the band at about -0.41 eV have a major contribution from the grain boundaries, as shown by the spatial distribution of the $\Gamma_2$ state (see Fig. 6(c)). In addition, the $\Gamma_2$ state shows quasi-localized decay away from the grain boundary. The states in the two bands crossing at the K state are extended and have almost equal distribution over all carbon atoms, as shown in Fig. 6(e). Our spin-polarized calculation also indicates that such 1D defect did not show any magnetic moment, which suggests that the 1D defect is not the origin of ferromagnetism observed in graphite [22,37,38]. Very recently, the theoretical calculation indicates that grain boundary dislocation of graphene with the core consisting of pentagon, octagon, and heptagon shows a small ferromagnetic contribution [39], which may be the origin of ferromagnetic behavior observed in Ref. 22.

## IV. CONCLUSIONS

To conclude, we present clear evidence for flat bands near Fermi energy in 1D extended defects on the surface of HOPG. The flat bands, which contribute to a pronounced peak in the tunnelling density of states, arise from $sp^2$ dangling bands (with antibonding nature) of undercoordinated carbon atoms at the edges of the defects. Our experimental result and analysis indicate that the presence of flat bands could be a universal behavior of similar 1D defects of graphene/graphite. The observed flat band in 1D extended defect can give rise to new correlated phases with desirable properties in graphene/graphite.

This work was supported by the National Natural Science Foundation of China (Grant Nos. 10721404, 10974019, 10974107, 11004010, 51172029 and 91121012) and the Fundamental Research Funds for the Central Universities.

§These authors contributed equally to this paper.
*Email: Junni@mail.tsinghua.edu.cn
helin@bnu.edu.cn.